# Improved Algorithms and Coupled Neutron-Photon Transport for Auto-Importance Sampling Method *

Xin Wang(王鑫)[1,2]  Zhen Wu(武祯)[3]  Rui Qiu(邱睿)[1,2;1)]  Chun-Yan Li(李春艳)[3]
Man-Chun Liang(梁漫春)[1]  Hui Zhang(张辉)[1,2]  Jun-Li Li(李君利)[1,2]

Zhi Gang(刚直)[4]  Hong Xu(徐红)[4]

[1] Department of Engineering Physics, Tsinghua University, Beijing 100084, China
[2] Key Laboratory of Particle & Radiation Imaging, Ministry of Education, Beijing 100084, China
[3] Nuctech Company Limited, Beijing 100084, China
[4] State Nuclear Hua Qing (Beijing) Nuclear Power Technology R&D Center Co. Ltd., Beijing 102209, China

**Abstract:** The Auto-Importance Sampling (AIS) method is a Monte Carlo variance reduction technique proposed for deep penetration problems, which can significantly improve computational efficiency without pre-calculations for importance distribution. However, the AIS method is only validated with several simple examples, and cannot be used for coupled neutron-photon transport. This paper presents improved algorithms for the AIS method, including particle transport, fictitious particle creation and adjustment, fictitious surface geometry, random number allocation and calculation of the estimated relative error. These improvements allow the AIS method to be applied to complicated deep penetration problems with complex geometry and multiple materials. A Completely coupled Neutron-Photon Auto-Importance Sampling (CNP-AIS) method is proposed to solve the deep penetration problems of coupled neutron-photon transport using the improved algorithms. The NUREG/CR-6115 PWR benchmark was calculated by using the methods of CNP-AIS, geometry splitting with Russian roulette and analog Monte Carlo, respectively. The calculation results of CNP-AIS are in good agreement with those of geometry splitting with Russian roulette and the benchmark solutions. The computational efficiency of CNP-AIS for both neutron and photon is much better than that of geometry splitting with Russian roulette in most cases, and increased by several orders of magnitude compared with that of the analog Monte Carlo.

## 1 Introduction

To solve deep penetration problems in radiation shielding calculation using Monte Carlo (MC) simulation, many methods have been developed based on the theory of coupling with deterministic method. These methods include different types of Monte Carlo variance reduction techniques [1,2,3,4] and coupled MC/discrete-ordinates methods [5,6,7]. However, these solutions have some limitations. The most effective Monte Carlo variance reduction techniques for deep penetration problems, for instance, importance sampling and geometry splitting with Russian roulette, require experience and pre-calculations for importance distribution, which are very time-consuming. Coupled MC/discrete-ordinates method must determine the interface position and convert the data structure between MC and discrete-ordinates simulations. MC and discrete-ordinates methods may be switched many times in some complicated deep penetration problems.

The Auto-Importance Sampling (AIS) method [8,9] is a new Monte Carlo variance reduction technique proposed by Tsinghua University for deep penetration problems, which can automatically adjust the particle importance distribution while transporting particles in layered space continuously. In general, the AIS method divides the whole geometry space into $K+1$ sub-spaces by introducing $K$ fictitious surfaces; particles are transported in these sub-spaces in sequence. The fictitious surface $k$ ($k=1,2…K$) is the Current Fictitious Surface (CFS) of sub-space $k$. In each sub-space, except for the last sub-space, fictitious particles are created on the CFS using next event estimators while transporting source particles and secondary particles. The source particles and secondary particles will be killed if they traverse the CFS. After all the source particles are simulated, the weights and number of fictitious particles are automatically adjusted using splitting/Russian roulette until the number of fictitious particles is as many as the source particles. Then, the fictitious particles are set as the source of the next sub-space, and the particle transport is performed in the next sub-space. More detailed information about the AIS method can be found in Ref. [8].

* Supported by the subject of National Science and Technology Major Project of China (2013ZX06002001-007, 2011ZX06004-007) and National Natural Science Foundation of China (11275110, 11375103)
1) E-mail: qiurui@tsinghua.edu.cn

Currently, the AIS method is implemented in MCNP5 [10] code, and can be applied to neutron or photon transport, separately. Compared with analog Monte Carlo, the computational efficiency of the AIS method is increased by about three orders of magnitude for several simple deep penetration problems. However, coupled neutron-photon transport calculations must be considered for some typical deep penetration problems in engineering, for instance, shielding calculations for reactors and high energy accelerators. Aside from this, the application of the AIS method is limited due to the following disadvantages:

1) The particles traversing the CFS during transport in each sub-space will affect the tally results around the CFS.

2) Fictitious particle creation lacks the capability of dealing with reflecting surfaces, white boundaries and periodic boundaries. The fictitious particle storage may overflow in the fictitious particles adjustment.

3) Only planar fictitious surfaces which are vertical to coordinate axes can be used.

4) The AIS method makes the random number stride overrun sometimes.

5) The calculation method of the estimated relative error is not precise enough.

Therefore, several novel improvements for the AIS method are described in this paper. Moreover, a completely coupled neutron-photon auto-importance sampling method is proposed with the improved algorithms. In this new method, the variance reduction techniques of the AIS method can be applied to neutrons and photons, simultaneously, in coupled neutron-photon Monte Carlo transport.

## 2 Materials and methods

### 2.1 Improved algorithms

#### 2.1.1 Particle transport

In the AIS method, when the particle travels in each sub-space, except for the last sub-space, the particle will be killed once it traverses the CFS, because its contributions to the current fictitious surface have been recorded by the fictitious particles. The particle position at each step of the random walk will be checked to find whether it is in the current or the next sub-space. Once the particle is located in the next sub-space, it will be killed. Although the fictitious particles can be created correctly in this way, the trajectory from the CFS to the position where the particle is killed is recorded redundantly. This will affect the tally results around the CFS, especially when a mesh tally is used.

In the improved particle transport algorithm, after a source or collision event, the distance to the CFS along the current direction will be calculated, and compared with the distance to the next geometry boundary and the next collision point. If the distance to the CFS is the minimum, it will be the distance of the next random walk, and the particle will be killed on the CFS.

As shown in Fig. 1, the particle will be killed at the next collision point A in the previous algorithm, and the trajectory $t$ is redundant. In the improved algorithm, $d$ will be the distance of the next random walk, and the particle will be killed at point B. Thus, it will ensure that the contributions to the whole geometry space are correct.

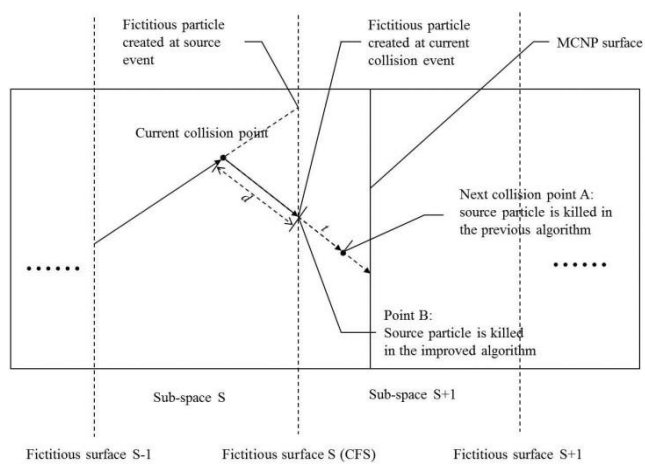

Fig.1. Improved particle transport algorithm

### 2.1.2 Fictitious particle creation and adjustment

In the AIS method, the fictitious particle weight is calculated by using next event estimators. The fictitious particle weight is equal to the probability that the particle has a collisionless free-flight to the CFS along the current direction after a source or collision event. The AIS method does not support reflecting surfaces, white boundaries and periodic boundaries, which leads to an underestimation of the fictitious particle weights and number. Actually, when the fictitious particle trajectory hits these surfaces or boundaries, its direction and location should be recalculated according to the type of the surface or boundary. Therefore, in the improved fictitious particle creation algorithm, fictitious particle creation methods dealing with reflecting surfaces, white boundaries and periodic boundaries are added to calculate the trajectory to the CFS and the fictitious particle location.

As mentioned above, splitting/Russian roulette is used in fictitious particle adjustment. When the weight of the fictitious particle is higher than the mean weight, the particle is split and stored in the fictitious particle storage. If the number of fictitious particles is larger than the number of source particles, the mean weight will be recalculated and redundant particles will be eliminated. In deep penetration problems with multiple materials, large changes may take place in the cross section data between different materials, which will make the weights of the fictitious particles fluctuate significantly. Additionally, when the small probability event that the source particle penetrates the shield is simulated, the weight of the resulting fictitious particle will be much higher than that of other fictitious particles, possibly many orders of magnitude higher. Hence, when the fictitious particles are split, a great number of "split fictitious particles" are generated, causing fictitious particle storage overflow. Considering that the "split fictitious particles" have exactly the same particle state, there is no need to store all of them in the fictitious particles storage. In the improved fictitious particles adjustment algorithm, a single unit of storage space is set to store the state of "split fictitious particles". Russian roulette is only performed on the serial numbers of fictitious particles, after which the fictitious particles are re-extracted according to the serial numbers.

In this way, storage space is saved, and the data overflow is avoided.

### 2.1.3 Fictitious surface geometry

The whole geometry space is divided into several sub-spaces by fictitious surfaces in the AIS method. Only planar fictitious surfaces which are vertical to the coordinate axes can be used. It cannot meet the demands in some deep penetration problems, for example, reactor pressure vessel neutron fluence calculations, which require cylindrical fictitious surfaces to divide the reactor. Therefore, cylindrical and spherical fictitious surfaces are added to the AIS method. Furthermore, rotation and translation operations of the fictitious surface are supported. The RDUM card of MCNP is used as the interface of the fictitious surface parameter setting. The fictitious surfaces and the MCNP surfaces are independent of each other. As a consequence, the AIS method is applicable to the problem with more complex geometry.

### 2.1.4 Random number allocation

The Russian roulette used in fictitious particle adjustment will cause a large consumption of random numbers. MCNP uses correlated sampling that the $i_{th}$ history will always start at the same point in the random number sequence. The value of the random number stride S allocated to each single history is only 152917 [10]. It cannot meet the demands of random numbers for fictitious particle adjustment, so S is exceeded sometimes.

Two random number sequences are used in the improved random number allocation algorithm. The first random number sequence RS1 is the one used in MCNP. In an AIS simulation, $K$ represents the number of fictitious surfaces and $N_{src}$ represents the number of source particles from source region. Then, the number of histories that needs to be calculated is $(K+1) \times N_{src}$. Similar to MCNP simulation, the $i^{th}$ history of the source particle from source region or fictitious surface will always start at the same point in the random number sequence RS1. Thus, $(K+1) \times N_{src}$ random number strides will be used for an AIS simulation. RS1 is only responsible for the normal random walk to avoid exceeding random number stride. The second random number sequence RS2 is used only for fictitious particle adjustment, and the initial random seed of RS2 is fixed. Different from MCNP simulation, RS2 is not segmented into many strides. Random numbers in RS2 are used one after another in an AIS simulation.

2.1.5 Calculation of the estimated relative error

The estimated relative error $R$ at the 1σ level in the AIS method is defined as Eq. (1) and Eq. (2) [11].

$$R = S_{\bar{x}} / \bar{x}. \quad (1)$$

$$S_{\bar{x}} = \left( \sum_{k=1}^{K+1} \frac{S_k^2}{N_k} \right)^{1/2}. \quad (2)$$

where $\bar{x}$ is the estimated mean, $S_{\bar{x}}$ is the estimated standard deviation of the mean $\bar{x}$, $N_k$ is the number of source particles in sub-space $k$ which is equal to $N_{src}$ in the AIS method, and $S_k$ is the estimated standard deviation of the mean weight of fictitious particles created in sub-space $k$ ($k=1,2...K$) or the estimated standard deviation of the mean $\bar{x}$ in the last sub-space.

The reason for using this error calculation method is that the AIS method uses a layered particle transport model, and $N_{src}$ source particles are transported in each sub-space. As a consequence, the contribution to the tally region of every single source particle from the source region cannot be obtained. However, this calculation method is not accurate. The states of the fictitious particles created on each fictitious surface are not able to accurately reflect the real distributions of weight, position, energy and angle; the error caused by particle elimination in fictitious particle adjustment is not recorded in $S_k$. Thus, in the improved algorithm of the estimated relative error calculation, the source particles are simulated by being divided into groups. $N_{src}$ source particles are divided into $m$ groups and $m$ simulations are performed. In each simulation, $N_{src}/m$ source particles are transported and the estimated mean of these $N_{src}/m$ particles $\bar{x}'$ can be calculated. Finally $m$ samples of $\bar{x}'$ are obtained, then the unbiased estimation of $S_{\bar{x}}$ is

$$S_{\bar{x}} = S_{\bar{x}'} / \sqrt{m}. \quad (3)$$

This algorithm is supposed to be more accurate since it accumulates all the transmission errors of the AIS method to the mean result of each group, avoiding calculating the errors of each sub-space in the AIS method.

## 2.2 Completely Coupled neutron-photon auto-importance sampling

In some radiation shielding designs, for instance, accelerator and reactor shielding, attenuation of neutrons and photons should be considered simultaneously. Accordingly, the deep penetration problems of coupled neutron-photon transport must be solved. In the AIS method, the variance reduction techniques can only be used for neutrons or photons separately. The fictitious surface should only be used for the particles of the same type as the source particles. Particles of a different type from the source particles are transported normally, and the calculation efficiency of these particles is not increased. In order to solve this problem, we propose a completely coupled Neutron-Photon Auto-Importance Sampling (CNP-AIS) method that can use the variance reduction techniques of the AIS method for both neutron and photon transport, simultaneously. In the CNP-AIS method, both neutron and photon fictitious surfaces are introduced. The geometry, type, and number/location of the fictitious surface should be determined:

A.  The geometry of the fictitious surface is chosen according to the geometry of the three-dimensional model and the tally region. For instance, planar fictitious surfaces are normally used for slab shielding problems; for reactor shielding problems, cylindrical fictitious surfaces are always used if the tally region is located at the lateral face of the pressure vessel.

B.  The fictitious surface type (neutron or photon) is chosen depending on the shielding effect of the material. If the shielding effect of the material is obvious for neutrons (or photons), neutron (or photon) fictitious surface should be introduced inside the material; if the material is suitable for both neutron and photon shielding, neutron and photon fictitious surfaces should be introduced simultaneously.

C.  In general, for a certain number of histories, more accurate results will be obtained with more fictitious surfaces, however, the simulation will cost more time. The optimum solution of the sub-space division is hard to determine. The sub-space division is relatively satisfactory when the penetrating probability is close to 1/10 in every sub-space [11].

The procedure of the CNP-AIS method is as follows:

1) From source region to tally region, a series of neutron and photon fictitious surfaces are introduced to divide the whole geometry space into several sub-spaces. The total number of neutron and photon fictitious surfaces is $K$, and the number of sub-spaces is $K+1$. The fictitious surface $k$ ($k=1,2…K$) is the CFS of sub-space $k$.

2) The Closest Photon Fictitious Surface (CPFS) and the Closest Neutron Fictitious Surface (CNFS) from the source are recorded.

3) The closest sub-space from the source is set to be the current sub-space, in which the particles will be transported, and the CFS is recorded. At least one of the CPFS and CNFS is the CFS. If the CPFS and CNFS are at the same location, they are both set to be the CFS.

4) The particles are transported from the source. At every source or collision event, fictitious particles are created on the CPFS or CNFS using next event estimators according to the particle type.

5) When the source particle or secondary particle arrives at the CFS, if the particle is a neutron (or photon) and the CNFS (or CPFS) is the CFS, the particle will be killed; if not, its state will be stored and transport will be stopped on the CFS. This ensures that all the neutrons and photons will not traverse the CFS.

6) After all the source particles are transported, the fictitious particles on the CFS will be adjusted to be as many as the source particles using splitting/Russian roulette. These fictitious particles, source particles and secondary particles stored on the CFS will be set as the source of the next sub-space. Then, the process will go back to step 1, and particle transport will be performed in the next sub-space.

When the closest fictitious surface of the same type as the particles is not the CFS, the reason why the particles are stored and stopped on the CFS in step 5 is shown in Fig. 2. If a neutron traverses the CFS (CPFS) in sub-space S after collision event 1 and has collision event 2 in which a secondary photon is generated, no photon fictitious particle will be created because the secondary photon is beyond the CPFS. A neutron transport in the CNP-AIS method is shown in Fig. 3.

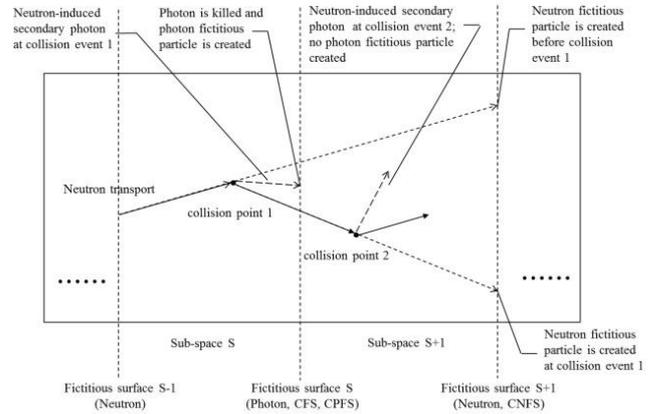

Fig.2. The situation when a particle traverses the CFS

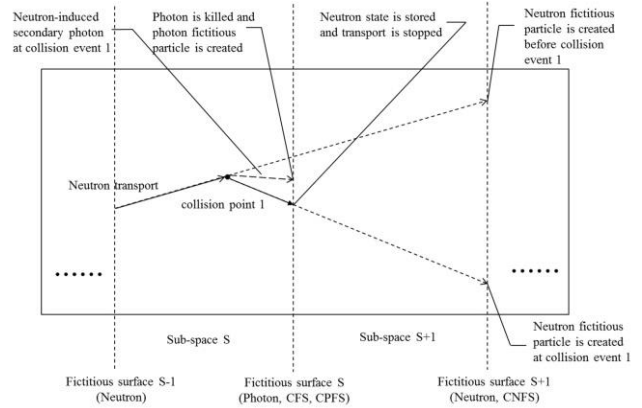

Fig.3. Particle transport in the CNP-AIS method

In the CNP-AIS method, except for the memory usage of analog Monte Carlo, additional memory should be allocated for the fictitious particles and source/secondary particles which are recorded on the fictitious surface. Because splitting/Russian roulette is used to adjust the number of fictitious particles, the space complexity of the CNP-AIS method is O(n), in which n represents the number of source particles.

Using these above variance reduction techniques of coupled neutron-photon transport and based on the improved AIS method presented in chapter 2.1, the CNP-AIS method was implemented in MCNP5 Code.

**2.3 Simulation set-up**

In order to validate the reliability of the CNP-AIS method, NUREG/CR-6115 PWR pressure vessel fluence calculation benchmark problems issued by the NRC [12] were calculated in this paper. The PWR model mainly consists of a 204 fuel assembly PWR core, a core barrel, thermal shield, vessel and an outer concrete biological shield. The power distribution is based on a detailed 15x15 fuel assembly pin-wise power distribution. The standard core loading pattern of the benchmark problems was used here. The azimuthal boundaries at 0 and 45 degrees were set to be reflecting boundaries. The outside of the biological shield wall, the top and bottom of the model were set to be void boundaries.

Five different examples including neutron/photon flux radial and axial distribution in the biological shield wall, neutron/photon cavity flux ($E>0.1$ MeV), neutron/photon flux spectrum at capsule location and neutron flux at pressure vessel 1/4 peak axial location ($E>1.0$ MeV) were calculated by using MCNP5 code with the methods of CNP-AIS, geometry splitting with Russian roulette (IMP-MC) and analog Monte Carlo (A-MC), respectively. The geometry importance distribution of neutrons was referred to the benchmark problem and the geometry importance distribution for photons was the same as that of neutrons.

In the NUREG/CR-6115 PWR benchmark, the thickness of the biological shield wall made of concrete is 213.36 cm, while the biological shield wall used in the benchmark calculation in Ref. [12] was only 45.085 cm thick. No calculations inside or outside the biological shield wall were performed. In the first two examples, the whole 213.36 cm biological shield wall was added to the PWR model. In the latter three examples, in order to allow consistent comparisons with DORT results, the PWR model used was the same as that used in Ref. [12], in which the biological shield wall was 45.085 cm thick.

A figure of merit (FOM) was used to evaluate the computational efficiency. The FOM is defined as:

$$FOM = 1/{R^2 \cdot T}. \quad (4)$$

All these calculations were performed on a notebook computer with Intel Core i7-3520M CPU 2.90 GHz and 16.0 GB memory.

## 3 Results and discussion

### 3.1 Neutron/photon flux radial distribution in biological shield wall

As shown in Fig. 4, 6 MCNP cylinder surfaces were added to biological shield wall. The fluxes of these 6 cylinder surfaces and the outer face of biological shield wall were tallied.

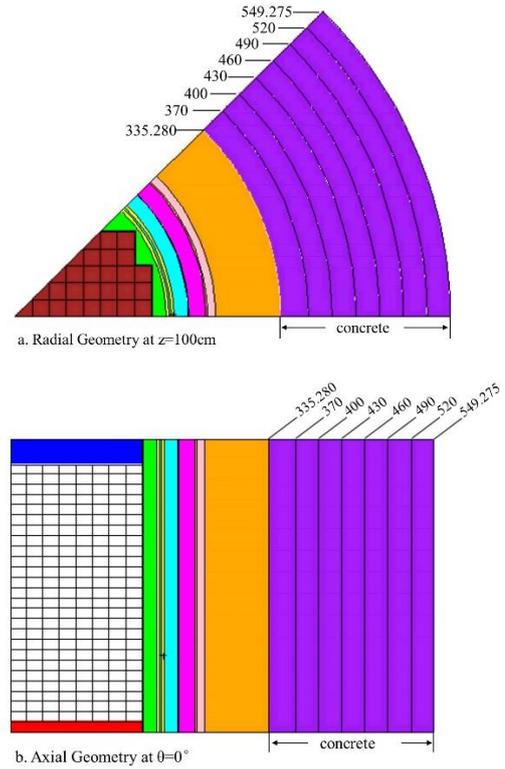

Fig.4. PWR model with full size biological shield wall (all dimensions in cm; the other parts of this model refer to Ref. [12]; color online)

In the IMP-MC simulation, the number of histories (NPS) was $4 \times 10^7$, and the computation time $T$ was 1905 minutes.

In the CNP-AIS simulation, eleven neutron and eleven photon cylindrical fictitious surfaces whose radii were 188, 215, 230, 340, 360, 390, 420, 450, 480, 510 and 530 cm, were introduced. NPS was $10^5$ and $T$ was 8 minutes.

The neutron and photon results are shown in Table 1 and Table 2 respectively. $R$ represents the estimated relative error. The neutron and photon FOM curves of IMP-MC and CNP-AIS are shown in Fig. 5 and Fig. 6 respectively.

Table 1. Neutron flux radial distribution in biological shield wall

| Surface No. | Radius (cm) | IMP-MC Flux (cm$^{-2}$·s$^{-1}$) | R | FOM (min$^{-1}$) | CNP-AIS Flux (cm$^{-2}$·s$^{-1}$) | R | FOM (min$^{-1}$) |
|---|---|---|---|---|---|---|---|
| 1 | 370 | 2.38E+09 | 0.003 | 7.20E+01 | 2.49E+09 | 0.046 | 6.01E+01 |
| 2 | 400 | 1.80E+08 | 0.006 | 1.74E+01 | 1.84E+08 | 0.045 | 6.15E+01 |
| 3 | 430 | 1.20E+07 | 0.018 | 1.62E+00 | 1.25E+07 | 0.051 | 4.90E+01 |
| 4 | 460 | 8.95E+05 | 0.066 | 1.20E-01 | 8.34E+05 | 0.053 | 4.43E+01 |
| 5 | 490 | 4.01E+04 | 0.218 | 1.10E-02 | 5.76E+04 | 0.063 | 3.19E+01 |
| 6 | 520 | 4.08E+03 | 0.735 | 9.72E-04 | 4.05E+03 | 0.071 | 2.50E+01 |
| 7 | 549.275 | - | - | - | 8.15E+01 | 0.087 | 1.66E+01 |

Table 2. Photon flux radial distribution in biological shield wall

| Surface No. | Radius (cm) | IMP-MC Flux (cm$^{-2}$·s$^{-1}$) | R | FOM (min$^{-1}$) | CNP-AIS Flux (cm$^{-2}$·s$^{-1}$) | R | FOM (min$^{-1}$) |
|---|---|---|---|---|---|---|---|
| 1 | 370 | 1.90E+09 | 0.003 | 7.77E+01 | 1.94E+09 | 0.038 | 8.66E+01 |
| 2 | 400 | 3.11E+08 | 0.005 | 2.59E+01 | 3.12E+08 | 0.039 | 8.30E+01 |
| 3 | 430 | 3.96E+07 | 0.010 | 4.95E+00 | 3.98E+07 | 0.037 | 9.23E+01 |
| 4 | 460 | 5.14E+06 | 0.027 | 6.99E-01 | 4.85E+06 | 0.045 | 6.12E+01 |
| 5 | 490 | 6.91E+05 | 0.065 | 1.25E-01 | 6.24E+05 | 0.031 | 1.33E+02 |
| 6 | 520 | 7.90E+04 | 0.189 | 1.48E-02 | 8.84E+04 | 0.024 | 2.13E+02 |
| 7 | 549.275 | 1.38E+04 | 0.250 | 8.42E-03 | 9.17E+03 | 0.026 | 1.88E+02 |

As shown in Table 1, for surface No. 1, 2, 3 and 4, the estimated relative errors of neutron results of IMP-MC and CNP-AIS were all below 10%, and the neutron results of IMP-MC and CNP-AIS were in good agreement. For surface No. 5, 6 and 7, IMP-MC could not give reliable results, whereas the estimated relative errors of neutron results of CNP-AIS were still below 10%. As shown in Fig. 5, the neutron FOM curve of IMP-MC had an exponential decay with the surface radius increasing, because the penetrating probability decreased through the concrete; however, the neutron FOM curve of CNP-AIS kept approximately stable. The neutron FOMs of CNP-AIS and IMP-MC for the first tally surface were almost equal, but the neutron FOM of CNP-AIS for the sixth tally surface was increased by four orders of magnitude compared with that of IMP-MC. The similar performance for photons can be seen in Table 2 and Fig. 6. In this example, CNP-AIS was much less time-consuming (only 8 minutes) than IMP-MC, but gave a much better performance.

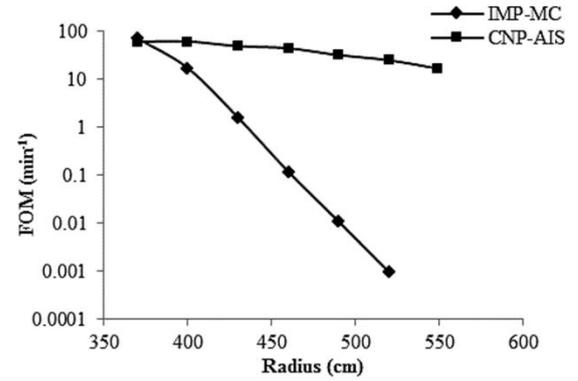

Fig.5. FOM curves of neutron flux radial distribution in biological shield wall

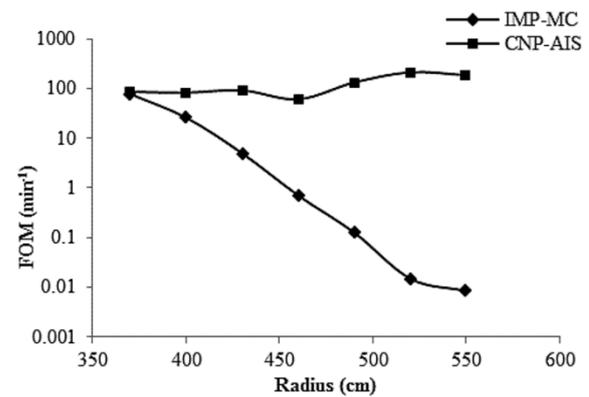

Fig.6. FOM curves of photon flux radial distribution in biological shield wall

### 3.2 Neutron/photon flux axial distribution in biological shield wall

The same PWR model as example 3.1 was used. Considering the computational efficiency of IMP-MC, an MCNP cylinder surface with radius of 490 cm was set to be the tally surface for neutrons, and an MCNP cylinder surface with radius of 520 cm was set to be the tally surface for photons. The tally surfaces were divided axially into 13 segments at the z values of 30, 60, 90, 120, 150, 180, 210, 240, 270, 300, 330 and 360 cm. The flux of each surface segment was tallied.

In the IMP-MC simulation, NPS was $4\times10^7$, and $T$ was 2112 minutes. The average estimated relative error of neutrons was 68.31% and that of photons was 55.82%.

In the CNP-AIS simulation, NPS was $4\times10^5$, and $T$ was 27 minutes. The fictitious surfaces introduced were the same as those of example 3.1. The average estimated relative error for neutrons was 5.38% and that for photons was 3.48%.

The results are shown in Fig. 7 and Fig. 8. The neutron and photon results of flux axial distribution in the biological shield wall of CNP-AIS were very accurate. However, IMP-MC could not give reliable results even with unacceptable time-consumption.

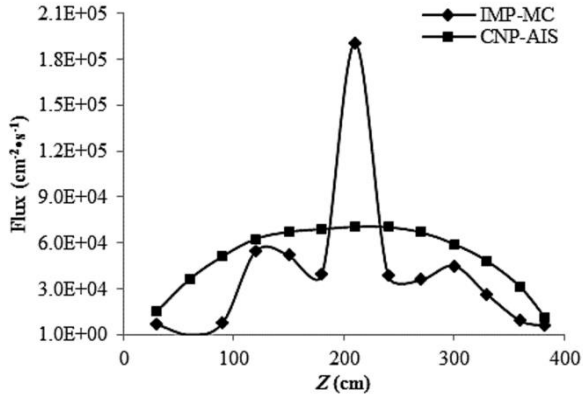

Fig.7. Neutron flux axial distribution in biological shield wall

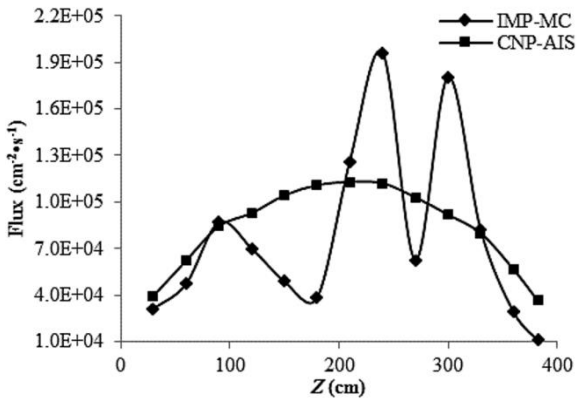

Fig.8. Photon flux axial distribution in biological shield wall

The neutron and photon FOM curves of CNP-AIS and IMP-MC for this example are shown in Fig. 9 and Fig. 10. The FOMs of CNP-AIS were increased by about four orders of magnitude compared with those of IMP-MC for both neutron and photon. CNP-AIS could give accurate results with reasonable time-consumption, whereas IMP-MC was not applicable to this example.

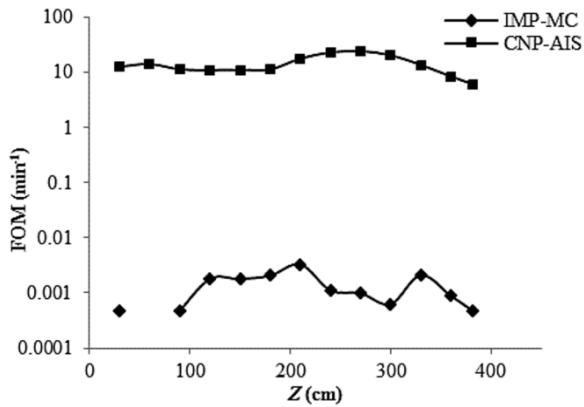

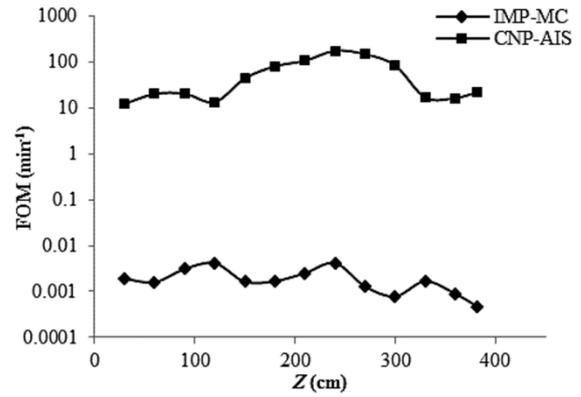

Fig.9. FOM curves of neutron flux axial distribution in biological shield wall

Fig.10. FOM curves of photon flux axial distribution in biological shield wall

### 3.3 Neutron/photon cavity flux ($E>0.1$ MeV)

In this example and the following examples, because no photon calculations were performed in Ref. [12], the comparison with DORT or MCNP4A was only available for neutron results. The tally regions were located at $r=320.06$ cm and $z=177.27$ cm, and distributed uniformly over 61 azimuthal locations.

In the IMP-MC simulation, NPS was $10^7$, and $T$ was 451 minutes. The average estimated relative error for neutrons was 3.12% and that for photons was 4.55%.

In the CNP-AIS simulation, NPS was $5\times 10^6$, and T was 221 minutes. Four neutron and four photon cylindrical fictitious surfaces, whose radii were 188, 215, 230 and 300 cm, were introduced. The average estimated relative error for neutrons was 1.78% and that for photons was 2.58%.

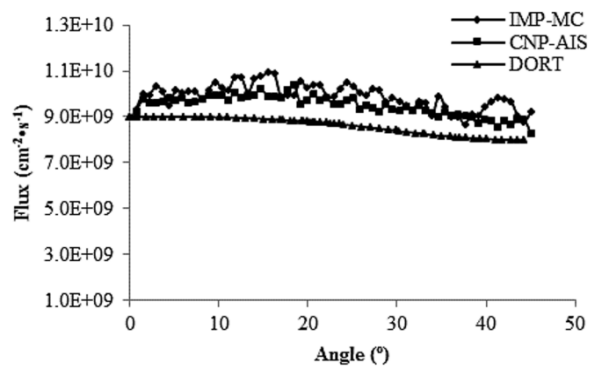

Fig.11. Neutron cavity flux ($E>0.1$ MeV)

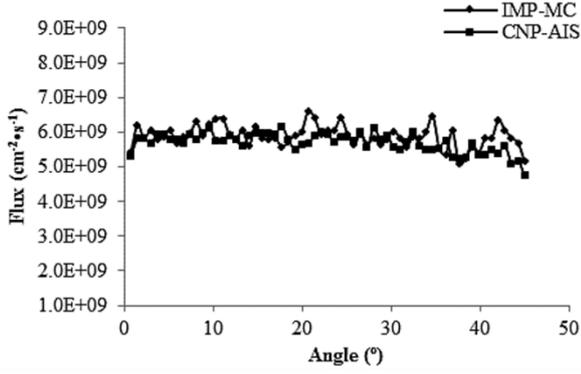

Fig.12. Photon cavity flux ($E$>0.1 MeV)

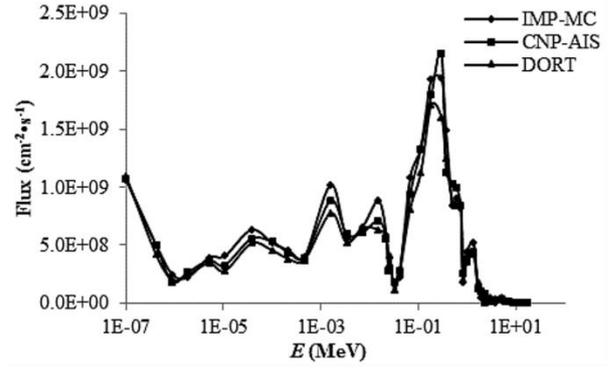

Fig.13. Neutron flux spectrum at capsule location

The DORT results in Ref. [12], and the calculation results of CNP-AIS and IMP-MC, are shown in Fig. 11 and Fig. 12. The results of IMP-MC and CNP-AIS are in good agreement. The average relative error for neutrons compared between IMP-MC and CNP-AIS is 4.66% and that for photons is 4.89%. Because of the different cross section data and calculation methods of DORT, its results are lower than the results of IMP-MC and CNP-AIS.

The neutron average FOM of IMP-MC was 2.28 min$^{-1}$ and that of CNP-AIS was 16.49 min$^{-1}$. The photon average FOM of IMP-MC was 1.07 min$^{-1}$ and that of CNP-AIS was 7.99 min$^{-1}$. In this example, computational efficiency increases of about eight times were observed by using CNP-AIS.

### 3.4 Neutron/photon flux spectrum at capsule location

The tally region was located at $r$=320.06 cm, $z$=177.27 cm and $\theta$=9.5°, and the 47-group energy structure was used for spectrum tally, which were the same as the DORT calculations in Ref. [12]. NPS, the setting of fictitious surfaces and $T$ were the same as those of example 3.3.

The DORT results, and the calculation results of CNP-AIS and IMP-MC, are shown in Fig. 13 and Fig. 14.

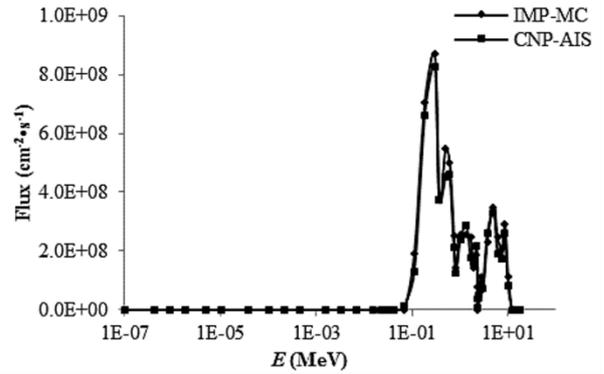

Fig.14. Photon flux spectrum at capsule location

In the IMP-MC simulation, the average estimated relative error of neutrons was 28.77% and that of photons was 26.94%. In the CNP-AIS simulation, the average estimated relative error of neutrons was 17.76% and that of photons was 18.77%. Referring to Fig. 13, although the neutron results of IMP-MC and CNP-AIS are not very accurate, they are in very good agreement with the results of DORT, considering that DORT results should be lower than Monte Carlo results. Referring to Fig. 14, the photon results of CNP-AIS are in good agreement with those of IMP-MC according to the estimated relative errors.

The neutron average FOM of IMP-MC was 0.13 min$^{-1}$ and that of CNP-AIS was 1.29 min$^{-1}$. The photon average FOM of IMP-MC was 0.05 min$^{-1}$ and that of CNP-AIS was 0.43 min$^{-1}$. In this example, computational efficiency increases of about one order of magnitude were observed by using CNP-AIS.

### 3.5 Neutron flux at pressure vessel 1/4 peak axial location ($E$>1.0 MeV)

The tally regions were located at $r$= 224.473cm and $z$=125.488 cm, and distributed uniformly over 20

azimuthal locations, which were the same as the MCNP4A calculations in Ref. [12].

In the IMP-MC simulation, NPS was $10^7$, and $T$ was 380 minutes. The average estimated relative error of the results was 4.18%.

In the CNP-AIS simulation, NPS was $10^7$, and $T$ was 314 minutes. Three neutron cylindrical fictitious surfaces, whose radii were 188, 208 and 222 cm, were introduced. The average estimated relative error of the results was 4.91%.

In the A-MC simulation, NPS was $10^9$, and $T$ was 2864 minutes. The average estimated relative error of the results was 20.59%.

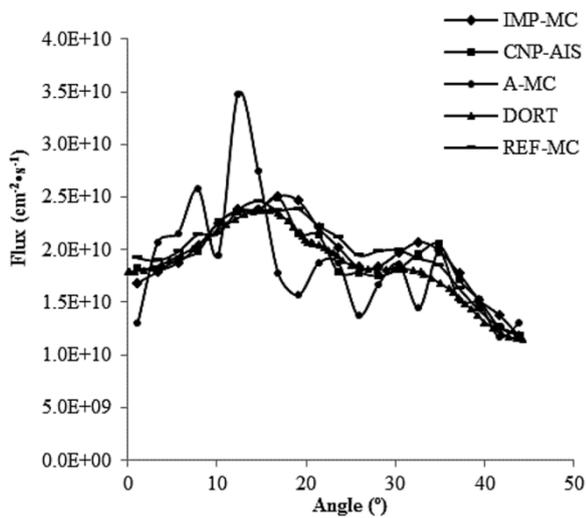

Fig.15. Neutron flux at pressure vessel 1/4 peak axial location ($E$>1.0 MeV)

The results of MCNP4A (REF-MC) and DORT in Ref. [12], and the calculation results of CNP-AIS, A-MC and IMP-MC, are shown in Fig. 15. The results of REF-MC, IMP-MC and CNP-AIS are in good agreement. The average relative error compared between CNP-AIS and IMP-MC is 3.92%, and that between CNP-AIS and REF-MC is 6.00%. Accurate results could not be obtained using A-MC without any variance reduction techniques within an acceptable period of time.

The FOM curves of CNP-AIS, IMP-MC and A-MC are shown in Fig. 16. The FOMs of CNP-AIS and IMP-MC are both increased by more than two orders of magnitude compared with those of A-MC. CNP-AIS has 10 FOMs higher and 10 FOMs lower than IMP-MC.

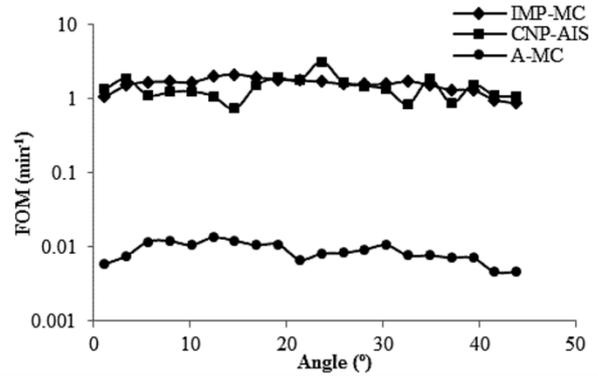

Fig.16. FOM curves of neutron flux at pressure vessel 1/4 peak axial location ($E$>1.0 MeV)

In this example, the performance of CNP-AIS was slightly inferior to that of IMP-MC, because the penetrating probability of pressure vessel 1/4 peak axial location was not very low, and CNP-AIS was designed to solve deep penetration problems. However, CNP-AIS could still give accurate results with very high computational efficiency.

## 4 Conclusions

In this paper, several improved algorithms for the AIS method are presented, and a completely coupled neutron-photon auto-importance sampling method is proposed. The CNP-AIS method was validated by the NUREG/CR-6115 PWR pressure vessel fluence calculation benchmark. The results showed that CNP-AIS method is applicable to different deep penetration problems, and improved the precision and efficiency of the Monte Carlo method.

The computational efficiency of the CNP-AIS method were much higher than that of geometry splitting with Russian roulette in all the examples, except for the neutron flux at pressure vessel 1/4 peak axial location example. In the neutron flux at pressure vessel 1/4 peak axial location example, compared with the analog Monte Carlo, the computational efficiency of the CNP-AIS method was increased by about two orders of magnitude. Therefore, it can be deduced that the computational efficiency of the CNP-AIS method would be even higher than that of the analog Monte Carlo in other examples. Improvement of computational efficiency became more obvious when the penetrating probability decreased. In the examples of flux radial and axial distributions in biological shield wall, geometry splitting with Russian roulette was not able to give reliable results within an acceptable period of time, while the CNP-AIS method could obtain very high calculation precision and efficiency. Furthermore, the CNP-AIS method simultaneously improved the computational efficiencies of both neutron and photon.

According to the results analysis and the comparison above, the validity of the CNP-AIS method for complicated deep penetration problems of coupled neutron-photon transport is proved. The CNP-AIS method can provide a reliable and efficient solution for deep penetration problems.